\documentclass[pdflatex,sn-mathphys-num,iicol]{sn-jnl}

\usepackage{graphicx}%
\usepackage{multirow}%
\usepackage{amsmath,amssymb,amsfonts}%
\usepackage{amsthm}%
\usepackage{mathrsfs}%
\usepackage[title]{appendix}%
\usepackage{xcolor}%
\usepackage{textcomp}%
\usepackage{manyfoot}%
\usepackage{booktabs}%
\usepackage{algorithm}%
\usepackage{algorithmicx}%
\usepackage{algpseudocode}%
\usepackage{listings}%
\usepackage{siunitx}
\usepackage{gensymb}

\usepackage{lineno}

\theoremstyle{thmstyleone}%
\theoremstyle{thmstyletwo}%

\theoremstyle{thmstylethree}%

\begin{document}

\title{Lagrangian dynamics unveil polymer conformation in viscoelastic flows}


\author[1,2]{\fnm{Louison} \sur{Thorens}}

\author[3,4]{\fnm{Gabriel} \sur{Juarez}}

\author[1,*]{\fnm{Jeffrey S.} \sur{Guasto}}

\author[4]{\fnm{Paulo E.} \sur{Arratia}}

\affil[1]{\orgdiv{Department of Mechanical Engineering}, \orgname{Tufts University}, \orgaddress{\city{Medford}, \postcode{02155}, \state{MA}, \country{USA}}}

\affil[2]{\orgdiv{Institute for Mechanobiology, Department of Bioengineering, College of Engineering}, \orgname{Northeastern University}, \orgaddress{\city{Boston}, \postcode{02115}, \state{MA}, \country{USA}}}

\affil[3]{\orgdiv{Department of Mechanical Science and Engineering}, \orgname{University of Illinois  Urbana-Champaign}, \orgaddress{\city{Urbana}, \postcode{61801}, \state{IL}, \country{USA}}}

\affil[4]{\orgdiv{Department of Mechanical Engineering and Applied Mechanics}, \orgname{University of Pennsylvania}, \orgaddress{\city{Philadelphia}, \postcode{19104}, \state{PA}, \country{USA}}}

\affil[*]{Corresponding author: Jeffrey.Guasto@tufts.edu}


\abstract{
The complex flow behavior of viscoelastic fluids emerges from a feedback between fluid stress and the out-of-equilibrium conformational dynamics of their constituent polymer chains.
While constitutive models can predict flow and stress fields in simple geometries, a fundamental understanding of the mechanisms regulating viscoelastic flows with mixed kinematics and non-trivial polymer dynamics remains elusive.
Quantifying the history-dependent conformation of polymer chains in flow is essential to resolve these complex systems.
To address this knowledge gap, we employ direct Lagrangian tracking of individual DNA molecules to reveal their conformational dynamics in microfluidic viscoelastic flows for different polymer contour lengths and concentrations. 
Comparing the measured molecular extension and orientation reveals discrepancies with canonical constitutive models, which do not fully capture the transient dynamics.
Our measurements show a distinct anisotropy in polymer relaxation and shape, which couple the polymer's rotation and extension through hydrodynamic drag, regulating their Lagrangian mechanics. 
These findings emphasize the importance of Lagrangian polymer history and the need to account for non-trivial intra-molecular mechanics to improve constitutive descriptions of viscoelastic flows.
}

\maketitle

The non-equilibrium dynamics of complex fluids emerge from molecular-scale processes, yet govern macroscopic phenomena across a multitude of natural systems, ranging from mucus transport and blood flow \cite{Stoodley2005,Datta2022} to landslides and glacier mechanics \cite{kostynick2022rheology,cuffey2010physics,coussot2017mudflow}. 
Elucidating the underlying mechanics of these processes are also critical to our ability to engineer new materials \cite{Larson_1999,de1993physics}, predict natural hazards \cite{coussot1996recognition,kostynick2022rheology}, and optimize industrial applications from polymer processing to pharmaceuticals \cite{Groisman2000,Virk1975,Burghelea2004,Arratia2006}.
In the case of polymeric solutions, their viscoelastic stress distribution and flow behavior are dictated by the ensemble-averaged flow kinematics experienced by their constituent polymer chains \cite{Bird1987,Larson1988}. 
The Lagrangian paths of individual polymers reflect spatio-temporal variations in shear, rotation, and extension, accumulating a complex deformation history that is not captured by simple, instantaneous descriptions \cite{Haward2012,Wagner2016}. 
Resolving the emergent properties of viscoelastic flows remains a challenge, particularly in mixed flow kinematics where existing constitutive models are only semi-quantitative \cite{Boyko2024}.

Continuum descriptions of macroscopic flows rely on constitutive models that coarse-grain complex polymer dynamics into a low-order conformation tensor \cite{Oldroyd1950,ByronBird1974,Bird1987,Doi1986,Boyko2024,Peterlin1966,sanchez2022understanding,Larson1988}.
These approaches model the evolution of the ensemble-averaged conformation, where heterogeneous chain structures are represented by a single degree of internal freedom and associated relaxation time. 
Such models thus commonly overlook the impact of polymer conformation and finite size on hydrodynamic drag and resulting orientational dynamics \cite{Boyko2024}.
Results from constitutive models are often compared to birefringence measurements \cite{Fuller1995}, which inherently average over microscopic heterogeneity, or simulations that employ the same mean-field assumptions \cite{Tirtaatmadja1995}. 
Consequently, the ensemble-averaged conformation used in mean-field approaches masks the complexity of single polymers, at the origin of ``molecular individualism'' \cite{deGennes1997,Perkins1997,Smith1999}.

Direct, single-molecule visualization has been integral to our understanding of polymer physics by revealing the non-equilibrium dynamics of individual polymers \cite{Perkins1997,Schroeder2003,Smith1999,Teixeira2004,chen2004conformation}.
Foundational works \cite{Perkins1997, Smith1999} showed that in homogeneous, Newtonian flows (i.e., purely extensional or shear), identical DNA chains unravel via distinctly different pathways, highlighting the critical role of their initial molecular configuration.   
These studies illustrate important limitations of constitutive models' mean-field approximation, which cannot fully capture nontrivial polymer topologies and hydrodynamics that regulate stretching \cite{Perkins1997}. 
However, polymer conformation evolves not only from the molecule's initial configuration but also from its history-dependent flow kinematics and the surrounding solvent rheology. 
Elucidating the mechanistic link between single-molecule dynamics and macroscopic flow behavior thus requires a Lagrangian understanding of individual polymer conformation in a non-Newtonian flow with spatially varying kinematics.

In this work, high-speed imaging of single DNA molecules is used to quantify the Lagrangian evolution of the polymer conformation as individual molecules traverse a spatially-varying, viscoelastic flow.
The resulting ensemble-averaged conformations provide spatially-resolved molecular extension and orientation fields that reveal a persistent discrepancy with canonical viscoelastic models. 
The observed discrepancy is traced to a distinct anisotropy in polymer relaxation and shape, in which the transverse and longitudinal axes relax at different rates. 
These anisotropic shape dynamics thus couple the polymer's rotation and extension through hydrodynamic drag, regulating their Lagrangian mechanics. 
These findings provide a first-principles physical basis, grounded in direct microscopic observation, to accurately capture transient dynamics in viscoelastic constitutive models.

\section*{Individual Lagrangian polymer dynamics in viscoelastic flow}

\begin{figure*}[t]
\centering
\includegraphics[width=2\columnwidth]{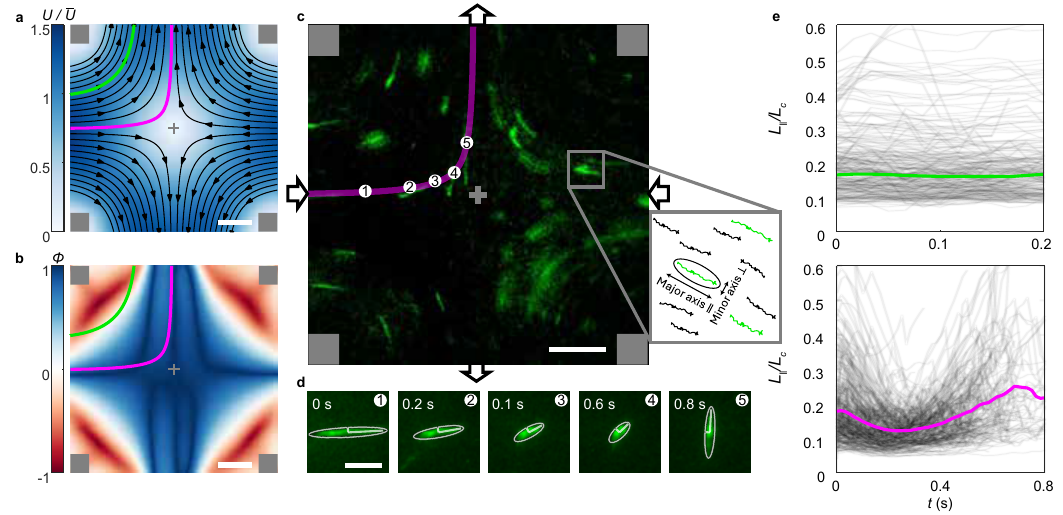}
\caption{\textbf{Lagrangian tracking of individual polymers in mixed kinematics viscoelastic flow.}
\textbf{a}. Steady, Newtonian base flow field (92\% w/w glycerol) quantified via particle tracking velocimetry in a microfluidic cross-slot ($W = H = \SI{100}{\micro\meter}$). Time-averaged flow speed ($U$) is normalized by the mean inlet speed ($\bar{U} = \SI{166}{\micro\meter}/\text{s}$). Grey cross and squares are hyperbolic point and channel corners, respectively. Highlighted streamlines (green, magenta) are consistent across panels. Scale bar, $\SI{20}{\micro\meter}$.
\textbf{b}. Flow type parameter ($\Phi$) of the base flow, distinguishing extension- ($\Phi\approx 1$) and shear-dominated ($\Phi\approx 0$) regions. Scale bar, $\SI{20}{\micro\meter}$.
\textbf{c}. Fluorescence microscopy of individual T4-DNA molecules ($c = c^*/8$; $\text{Wi} = 0.12$). Scale bar, $\SI{20}{\micro\meter}$. Inset: Schematic of DNA suspension with a fluorescently labelled subpopulation. 
\textbf{d}. Time sequence of a single molecule (magenta streamline in \textbf{c}), which enters pre-stretched, undergoes relaxation and compression, then rotates and stretches along the extensional flow axis. White ellipses represent the measured radius of gyration tensor. Scale bar, $\SI{10}{\micro\meter}$.
\textbf{e}. Lagrangian extension ($L_\parallel/L_c$) of molecules along shear- (upper) and extension-dominated (lower) streamlines (green and magenta, respectively, in \textbf{a}-\textbf{b}). Grey lines are individual polymers, and colored lines are ensemble averages.
}
\label{fig1}
\end{figure*}

The Lagrangian dynamics of individual polymers are characterized in a steady microfluidic cross-slot flow (Fig.~\ref{fig1}a). The microchannel has a square cross-section with equal height and width ($H=W = \SI{100}{\micro\meter}$), and the Newtonian base flow field exhibits mixed kinematics (Fig.~\ref{fig1}b) as quantified by the flow-type parameter, $\Phi$. 
The near-wall regions are dominated by rotational ($\Phi \approx -1$) and shear ($\Phi \approx 0$) components, while the central region and stagnation point are dominated by extension ($\Phi \approx 1$; see Methods for details).  
Different polymeric solutions are prepared by suspending DNA molecules having different contour lengths ($L_c$) and concentrations ($c$) in a Newtonian solvent (92\% w/w glycerol). 
Two double-stranded DNA molecules, $\lambda$-phage ($L_c \approx \SI{16.5}{\micro\meter}$) and T4 ($L_c \approx \SI{55.6}{\micro\meter}$) were investigated in the dilute ($c/c^*=1/8$; T4-DNA) and semi-dilute ($c/c^*=1$; $\lambda$- and T4-DNA) regimes, where $c^*$ is the overlap concentration. 
A small fraction (parts-per-billion, ppb) of the molecules in solution are fluorescently labelled \cite{Kawale2017}, enabling high-fidelity Lagrangian tracking of individual polymer conformations (see Methods; Fig.~\ref{fig1}c).

Experiments are performed by injecting the polymer solution into the microchannel inlets at a prescribed flow rate (mean inlet flow speed, $\bar{U}$). 
Under dilute conditions ($c=c^\ast/8$; T4-DNA), the polymeric flow is weakly non-Newtonian, characterized by a Weissenberg number, $\textrm{Wi} = \tau\dot{\epsilon} \approx 0.12$, where $\tau$ is the fluid relaxation time as measured by extensional rheology (see Methods), and $\dot{\epsilon} = \bar{U}/W$ is the strain rate. 
The instantaneous shape and orientation of individual molecules ($k$) are quantified by their intensity distribution via the radius of gyration tensor, $\mathbf{G}_k(t)$, as an established proxy for their conformation (see Methods; Fig.~\ref{fig1}d; Supplementay Video 1) \cite{Teixeira2004}. The fractional extension, $L_\parallel/L_c$, is defined using the maximum eigenvalue of $\mathbf{G}_k$ (see Supplementary Information).

Individual polymers exhibit highly dynamic conformations as they are advected through space and time (Fig.~\ref{fig1}c-d). While all molecules are consistently pre-stretched at the cross-slot entrance (Fig.~\ref{fig1}e), the temporal evolution of their extension is dictated by their local flow history and varies dramatically across different molecules and different regions of the channel. 
Molecules flowing near the channel walls are advected by shear-dominated flow (Fig.~\ref{fig1}b), where the ensemble average conformation maintains an approximately constant stretch with no net change in extension or contraction (Fig.~\ref{fig1}e, upper). 
In contrast, polymers flowing near the hyperbolic stagnation point (Fig.~\ref{fig1}b) are subjected to strong extension and undergo a distinct stretch-coil-stretch transition (Fig.~\ref{fig1}e, lower), which markedly deviates from the uniform coil-stretch unfolding predicted by constitutive model simulations~\cite{Kumar2023}. 
In both cases, individual molecules advected along similar streamlines exhibit considerable heterogeneity in their extension.
These observations demonstrate that in complex flows, molecular individualism acquires a Lagrangian character, whereby identical molecules exhibit divergent conformational evolution set by their unique initial polymer chain topology and flow history.

\section*{From Lagrangian to Eulerian polymer conformation}

\begin{figure*}[t]
\centering
\includegraphics[width=2\columnwidth]{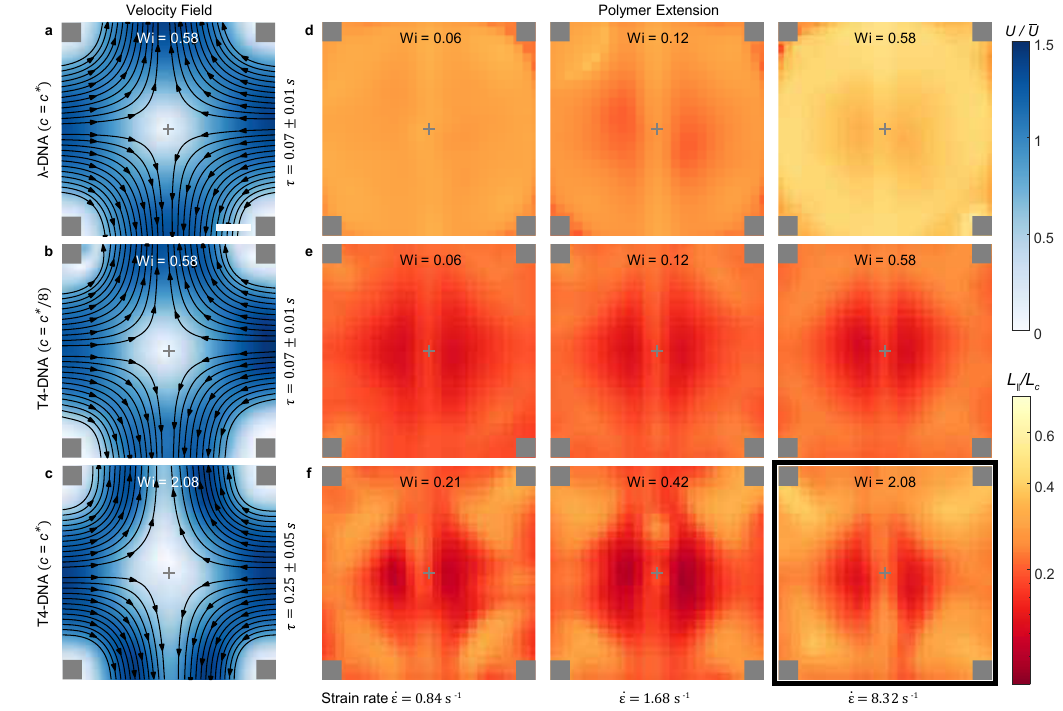}
\caption{\textbf{Polymer concentration and contour length impact flow topology and polymer extension across Weissenberg numbers.}
\textbf{a-c}. Time-averaged flow speed fields with streamlines at the highest $\text{Wi}$ tested for each fluid and the same imposed strain rate ($\dot{\epsilon} = 8.32$~s$^{-1}$). Only T4-DNA ($c = c^*$) exhibits noticeably non-Newtonian flow behavior. Scale bar, $\SI{20}{\micro\meter}$.
\textbf{d-f}. Spatially-resolved, ensemble-averaged Eulerian polymer extension maps ($L_\parallel/L_c$) determined from time-resolved Lagrangian conformation data (see Fig.~1).}
\label{fig2}
\end{figure*}

To quantify the impact of individual molecular stretching in non-Newtonian flows, Eulerian conformation fields are constructed from the experimentally measured Lagrangian polymer conformations across different molecules, concentrations, and flow conditions. 
Leveraging robust single-molecule statistics, Eulerian maps of polymer extension (Fig.~\ref{fig2}), $L_{||}/L_c$, are obtained by locally averaging the molecular conformation from individual time-resolved Lagrangian trajectories, $\mathbf{x}_k(t)$, in these steady flows. Here, the local mean radius of gyration tensor \cite{Schroeder2003}, $\mathbf{G}(\mathbf{x}) = \langle \mathbf{G}_k(\mathbf{x}(t)) \rangle_{k,\mathbf{x}\in \pm \delta \mathbf{x}}$, is computed as the average over all molecules whose center-of-mass resides within a local neighborhood ($\delta \mathbf{x}$) of an Eulerian grid. 
Polymer solution relaxation times vary with both contour length and concentration (Fig.~\ref{fig2}). 
Through precise control of the flow rate, Weissenberg numbers ($\text{Wi}$) span the weakly viscoelastic regime to well above the expected coil-stretch transition ($\text{Wi} \in [0.06, 2.08]$).  

Measured, time-averaged velocity fields (Fig.~\ref{fig2}a-b) of semi-dilute $\lambda$-DNA ($c = c^*$) and dilute T4-DNA ($c = c^*/8$) are weakly viscoelastic flows (Wi=0.58), with flow topologies comparable to the Newtonian base flow (Fig.~\ref{fig1}a). In these cases, the molecular extension maps exhibit similar topologies across $\text{Wi}$  (Fig.~\ref{fig2}d-e): 
Molecular extension decreases when approaching the hyperbolic point (stretch-coil transition), followed by significant stretching along the extensional direction (coil-stretch transition), and finally reaching approximately $50\%$ of full extension. 

A marked departure from Newtonian flow emerges for the semi-dilute T4-DNA suspension ($c = c^*$; Fig.~\ref{fig2}c), which is absent from all $\lambda$-DNA and dilute T4-DNA cases (Fig.~\ref{fig2}a-b, Fig.~\ref{fig:SI_velocityFields}).  
An extended low-velocity region develops immediately downstream of the hyperbolic point, a flow structure found in viscoelastic cross-slot flows \cite{Yokokoji2023-ux}.
Here, the emergence of the low velocity zone is accompanied by a pronounced region of high polymer stretching along the extensional manifold of the flow (Fig.~\ref{fig2}f). This non-Newtonian flow topology is triggered even at relatively low Weissenberg number ($\text{Wi} = 0.21$) for semi-dilute T4-DNA (Fig.~\ref{fig:SI_velocityFields}). We note, however, that the quoted $\text{Wi}$ values are likely underestimated by extensional rheology \cite{Sousa2015}, consistent with substantially longer relaxation times from single-molecule measurements (Fig.~\ref{fig4}). 
Despite the non-Newtonian response of the flow field, the maximal polymer extension remains comparable to the lower $\text{Wi}$ cases (Fig.~\ref{fig2}d-f). 
These results are the first direct experimental reconstruction of spatially-resolved Eulerian polymer conformation fields derived from Lagrangian histories of individual molecules, providing a benchmark for evaluating classical constitutive models.

\section*{Comparison with constitutive models}

\begin{figure*}[t]
\centering
\includegraphics[width=2\columnwidth]{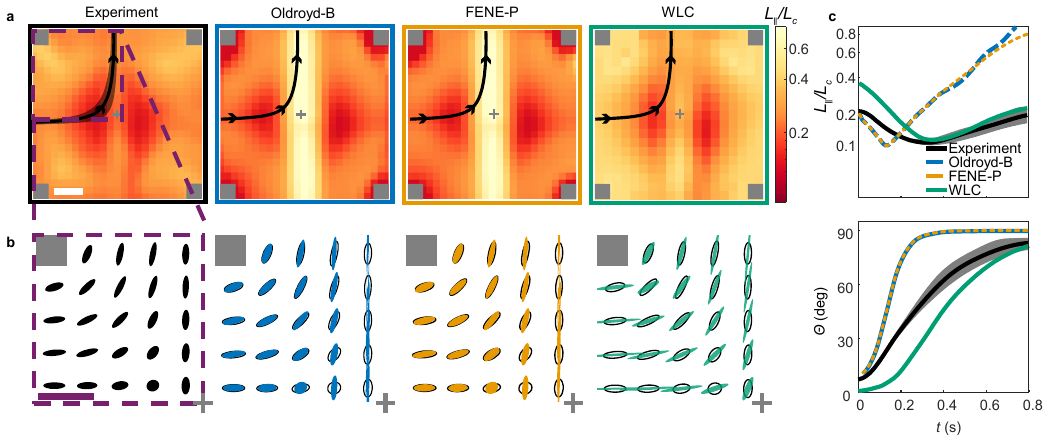}
\caption{\textbf{Comparison of experimentally measured polymer extension with constitutive and molecular models.}
\textbf{a}. Measured polymer extension (T4-DNA for $c=c^*$ and $\text{Wi} = 2.08$; see also Fig.~\ref{fig2}f) compared to Oldroyd-B, FENE-P, and WLC model predictions. Models utilize the measured flow field (Fig.~2c) to determine polymer deformation. Inlet conformation for Oldroyd-B and FENE-P are matched to experiments. Scale bar, $\SI{20}{\micro\meter}$
\textbf{b}. Visualization of the radius of gyration tensor for the quadrant shown in \textbf{a}. Black outlines are experimental data for comparison.  Scale bar, $\SI{20}{\micro\meter}$.
\textbf{c}. Lagrangian polymer extension (upper) and orientation (lower) as a function of time along a representative streamline (\textbf{a}, magenta).
Shaded area corresponds to the range of streamlines considered in \textbf{a}.
}
\label{fig3}
\end{figure*}

Experimentally measured molecular conformation fields enable comparison with a hierarchy of constitutive models, from continuum dumbbell descriptions to a resolved Worm-like Chain model (Fig.~\ref{fig3}).  
Viscoelastic constitutive models evolve the conformation tensor, $\mathbf{A} \equiv \langle \mathbf{R}\mathbf{R}^\top \rangle$, in flow, where $\mathbf{R}$ is a molecule's end-to-end vector \cite{Bird1987, Larson1988}. A direct comparison of the spatio-temporal conformation dynamics between the models and experiments is achieved by noting that the radius of gyration tensor, $\mathbf{G}$, shares the same principal orientation as $\mathbf{A}$, with their longitudinal extension remaining proportional (Fig.~\ref{fig:SI_conformation_WLC}) \cite{Larson1999}.
Focusing on the high Weissenberg number T4-DNA case ($\text{Wi} = 2.08$; $c/c^\ast = 1$), three different models are simulated (Fig.~\ref{fig3}a; see Methods and Supplementary Information): 
(\textit{i}) the Oldroyd-B model assumes an infinitely extensible Hookean dumbbell \cite{Oldroyd1950}; (\textit{ii}) the FENE-P model, while similar to the Oldroyd-B model, enforces finite spring extensibility (and thus polymer stress) \cite{Peterlin1966}; and (\textit{iii}) the Worm-like Chain (WLC) model is a discrete bead-spring chain that preserves high internal degrees of polymer freedom \cite{Marko1995, Larson1999}. Importantly, all three models are evolved using the same experimentally measured velocity field (Fig.~\ref{fig2}c) and have matched relaxation times. While the WLC chain conformation naturally recovers the observed pre-stretch in the inlet flow (see Methods), constitutive model conformations are initialized to match the experiments at the inlet.

All three models qualitatively reproduce the main features of the extension field topology relative to the experiment (Fig.~\ref{fig3}a-b). 
Near the wall, where the flow is shear- and rotation-dominated, the Oldroyd-B and FENE-P constitutive models agree relatively well with the experiments. 
However, these dumbbell models predict a significantly larger polymer stretching along the extensional manifold of the flow (Fig.~\ref{fig3}a). 
A closer examination of the full conformation tensor (Fig.~\ref{fig3}b) reveals that this observed hyper-extension is associated with an over-predicted preferential alignment near the extensional manifold. 
Notably, the finite extensibility imposed by the FENE-P model does not resolve these discrepancies. 
In contrast, the WLC simulations, which are known to capture complex polymer microstructure \cite{Larson1999,Hur2000}, more closely reproduce the topology of the measured extension maps (Fig.~\ref{fig3}a). 
Importantly, the WLC model only slightly overestimates the polymer extension and correctly preserves the measured orientation (Fig.~\ref{fig3}b; see also Fig.~\ref{fig:SI_orientationMaps}).
These results emphasize that resolving the polymer chains' internal degrees of freedom -- and thus finite spatial extent -- is a critical model feature.

\section*{Molecular complexity regulates transient conformation dynamics}
The Eulerian frame comparisons above establish the conditions under which constitutive models diverge from experiments. 
However, understanding the origin of this divergence requires examining the Lagrangian polymer history, where the transient molecular dynamics are directly accessible. 
Under steady-state conditions, polymer extension ($L_{||}/L_c$) and orientation ($\theta$) are measured in time (Fig.~\ref{fig3}c) along a representative near-hyperbolic streamline (Fig.~\ref{fig3}a, black curve).
Constitutive models correctly predict an initial compression phase but reach minimum extension $\approx 3.5$ times faster than observed in experiments ($\approx 0.1$~s versus $\approx 0.35$~s; Fig.~\ref{fig3}c, upper). 
Similarly, over-prediction of the extension persist into the stretching phase, where constitutive models are unaffected by finite extensibility, and the stretching rate is dictated strictly by $\text{Wi}$ \cite{Bird1987} (Fig.~\ref{fig3}c).
Importantly, this accelerated stretching is accompanied by a rapid reorientation along the extensional flow direction (Fig.~\ref{fig3}c, lower). 
On the other hand, the experiment and the WLC model exhibit both a slower molecular rotation and extension (Fig.~\ref{fig3}c), where the former is visualized in spatially resolved maps of Eulerian orientation (Fig.~\ref{fig:SI_orientationMaps}).
These Lagrangian measurements illustrate the importance of the kinematic coupling between polymer orientation and their rate of elongation in the extensional flow direction.

Our experimental and WLC results highlight an inherent limitation of many dumbbell-based models \cite{Boyko2024}, which represent a polymer with a single internal degree of freedom \cite{Larson_1999}. 
Notably, because the non-interacting, dilute WLC simulations reproduce experimental observations even at the overlap concentration ($c = c^*$), this limitation cannot be attributed to the lack of intermolecular interactions in simulations \cite{Stoltz2006,Hsiao2016}. 
Polymer chains with high internal degrees of freedom manifest as complex two- (or three-) dimensional objects.
The anisotropic shape of these polymers couples to fluid velocity gradients through conformation-dependent hydrodynamic drag \cite{deGennes1974, Hinch1994, Schroeder2003}.
Their evolving shape governs their rate of rotation \cite{Jeffery1922}, ultimately feeding back on their extension. 
To determine if this geometric coupling drives the observed transient delay in extension, the Lagrangian evolution of the polymer along its principal directions is examined independently.

\begin{figure}[h]
\centering
\includegraphics[width=\columnwidth]{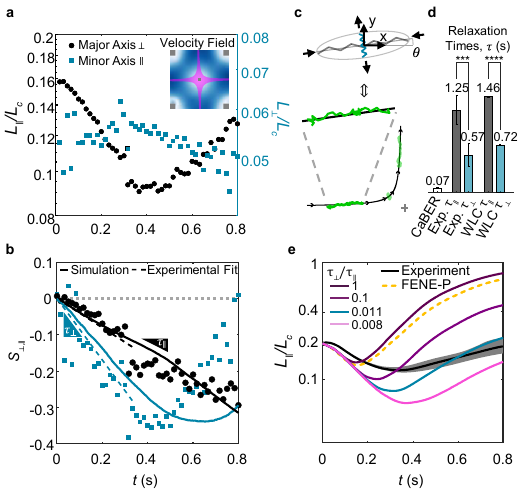}
\caption{\textbf{Single molecule mechanics reveal anisotropic polymer relaxation time.} 
\textbf{a}. Measured stretch-coil-stretch DNA dynamics (T4-DNA; $c = c^*/8$, $\text{Wi} = 0.06$) for the major and minor radius of gyration axes in primarily extensional flow. Inset: Region of analysis (magenta).
\textbf{b}. Residual deformation parameter, $S_{||,\perp}(t)$, isolates polymer relaxation dynamics from local fluid deformation (see Methods). Relaxation times measured from $S(t)$ slope (dashed lines) for experiments (markers) and WLC simulations (solid lines) show longer relaxation times along major compared to minor polymer axes. Error bars in \textbf{d} represent the 95\% confidence intervals of the linear fits.
\textbf{c}. Anisotropic polymer dynamics (and corresponding anisotropic dumbbell model) measured by relaxation times along major ($\tau_\parallel$, black) and minor ($\tau_\perp$, blue) axes to quantify observed deformation and orientation in lab frame ($\theta$).
\textbf{d}. Measured polymer relaxation times exhibit significant anisotropy (z-test) along major versus minor axes, which is consistent with WLC simulations but higher compared to capillary break-up extensional rheology (CaBER; see Methods).
\textbf{e}. Average polymer extension predicted by anisotropic dumbbell simulations for varying relaxation time ratios ($\tau_\parallel/\tau_\perp$; fixed $\tau_\parallel=\SI{1.25}{\second}$), where $\tau_\parallel/\tau_\perp = 0.011$ (blue) achieves best agreement with experiments (black). The shaded area corresponds to the range of streamlines considered (Fig.~\ref{fig3}a,c).}
\label{fig4}
\end{figure}

\section*{Anisotropic shape relaxation in stretched polymers}
To further elucidate the microscopic mechanisms regulating polymer conformation in flow, the anisotropic shape relaxation of polymers is examined. Operating in the dilute regime ($c = c^*/8$) to preclude any inter-chain hydrodynamic and entanglement effects, the Lagrangian polymer deformation is decomposed along its principal axes, $L_{||,\perp}$ (Fig.~\ref{fig4}a,c).
The total rate of polymer compression/extension, $\sim \dot{\epsilon}_{i}(t) + \tau_{i}^{-1}$, is due to both the instantaneous rate of fluid deformation and intrinsic relaxation along polymer axis, $i$.
To extract the molecule's principal relaxation times, $\tau_{||,\perp}$, we consider the residual deformation, $S_{||,\perp}$, which decouples the flow-induced polymer deformation using the local deformation gradient tensor (see Methods).
The longitudinal residual deformation decreases linearly with time as expected.
The transverse component decays more rapidly and deviates at later times (Fig.~\ref{fig4}b, symbols), corroborated by WLC simulations (Fig.~\ref{fig4}b, solid lines). 
The principle relaxation rates reveal a striking anisotropy, whereby the transverse relaxation time ($\tau_{\perp}$) is approximately half that of the longitudinal axis ($\tau_{\parallel}$; Fig.~\ref{fig4}d). 

To determine whether the observed anisotropic shape relaxation accounts for the discrepancy in polymer extensional dynamics (Fig.~\ref{fig3}), a phenomenological anisotropic FENE-P model is developed (Fig.~\ref{fig4}e; see Methods). 
The polymer conformation is represented by two finitely-extensible, orthogonal dumbbells that deform with the fluid gradient but relax at independent rates, $\tau_{\parallel,\perp}$ (Fig.~\ref{fig4}c). 
Considering the flow along the highlighted streamline in Fig.~\ref{fig3}a, setting both relaxation rates to the longitudinally measured rate ($\tau_\perp = \tau_\parallel = \SI{1.25}{\second}$) successfully recovers the standard isotropic FENE-P prediction (Fig.~~\ref{fig4}e).
However, a parametric exploration yields a best-fit transverse relaxation time of $\tau_\perp \approx 0.011 \tau_\parallel \approx \SI{13}{\milli\second}$. 
Although smaller than the measured $\tau_\perp$ (Fig.~\ref{fig4}d), this transverse time is remarkably consistent with the sub-chain relaxation time of the Rouse model ($\tau_0 \approx \SI{8}{\milli\second}$; see Supplementary Information) \cite{Rouse1953,Doi1986}.
This result suggests that longitudinal polymer extension is governed by the global chain unfolding ($\tau_{||}$), while transverse dynamics are limited by the rapid equilibration of sub-chains ($\tau_{\perp} \approx \tau_0$).

The observed anisotropic shape relaxation strongly couples molecular stretching to non-trivial flow kinematics through polymer rotation. 
Akin to Jeffery orbits for rigid ellipsoidal particles \cite{Jeffery1922}, the rate of alignment along the extensional flow direction depends on the polymer's effective hydrodynamic aspect ratio. 
Because flexible linear polymers maintain a finite two-dimensional footprint, they incur a rotational ``inefficiency'' in extensional flow alignment \cite{Boyko2024}, regulated by a rapid transverse relaxation ($\tau_\perp \ll \tau_\parallel$).
The resulting retardation of polymer alignment to the extensional axis thus suppresses the rapid coil-stretch transition predicted by dumbbell models and rationalizes the observed delay in elongation (Figs.~\ref{fig3}c~and~\ref{fig4}e).
Some continuum models (e.g., FENE-PTML \cite{PhanThien1984}) have proposed a stretch-dependent anisotropic drag to correct for rotational inefficiency, but these remain continuum-level approximations. 
The present work directly quantifies the impact of hindered molecular rotation in direct microscopic observations, and emphasizes the importance of anisotropic shape relaxation on the macroscopic conformation field.

\section*{Conclusion}
This work elucidates the critical link between microscopic molecular complexity and macroscopic continuum descriptions of viscoelastic flows. 
Through direct single-molecule imaging of polymer dynamics, these experimental results uniquely provide direct access to the Lagrangian conformational history of individual polymers advected through a non-Newtonian flow with spatially varying kinematics. 
From Lagrangian polymer dynamics, spatially-resolved Eulerian conformation fields were constructed and compared to classical constitutive models, revealing that the observed polymer stretch-coil-stretch transition near the hyperbolic point is significantly slower than theoretically predicted.
A Worm-like Chain model corroborates this discrepancy and highlights the need to account for internal degrees of polymer freedom.

Detailed examination shows that the rate of polymer stretching is strongly coupled to polymer orientation relative to the extensional flow direction. 
By decoupling the measured molecular deformation history in flow along principle polymer strain axes, anisotropic shape relaxation is revealed. 
Previous theoretical works established that polymer extension increases backbone drag \cite{Hinch1994,Schroeder2003}, consistent with conformation hysteresis in extensional flow \cite{deGennes1974}, and drives anisotropic mobility in concentrated solutions and melts \cite{Giesekus1982,PhanThien1984}. 
Our results reveal a complementary mechanism at the individual molecule level: dictated by the finite transverse size of linear flexible polymers, longitudinal and transverse deformations are governed by fundamentally different relaxation timescales.
Crucially, this structural anisotropy controls the hydrodynamic signature of the polymer (i.e., effective aspect ratio), which decreases their rate of rotational alignment toward the extensional flow direction and ultimately, their Lagrangian rate of stretching. 
Deriving classic constitutive models from idealized, zero-thickness dumbbells fundamentally neglects transverse hydrodynamic friction, yielding conformation tensors that omit this essential rotational retardation.

Our observations are consistent with a broader consensus that existing continuum descriptions of viscoelastic flow remain insufficient for flows with mixed or unsteady kinematics \cite{Datta2022}. 
A more comprehensive set of constitutive models incorporating finite extensibility, conformation-dependent drag, and non-affine deformation \cite{Boyko2024} remains largely untested against direct polymer conformation measurements.
The fundamental new observations from the present single molecule experiments provide the necessary insights for developing improved, physically-grounded viscoelastic constitutive models. 
To accurately predict fluid dynamic instabilities and transport properties in viscoelastic flows, data-driven constitutive models must account for the high internal degrees of freedom of real polymers and the accompanying complex shape dynamics.

\vspace{0.5in}
\noindent
\textbf{Data availability.}
The data that support the findings of this study are available via
Harvard Dataverse database [URL to be inserted].
\\

\noindent
\textbf{Code availability.}
The codes used for the analysis and simulations in this study are available via
Harvard Dataverse database [URL to be inserted].
\\

\noindent
\textbf{Acknowledgements.}
The authors thank M.D. Graham, T. Gao, and A.N. Morozov for scientific discussions. The work of J.S.G. and L.T. was supported by the U.S. National Science Foundation (NSF) through awards CMMI-2027410 and CBET-2141349, and through the Gordon and Betty Moore Foundation, grant DOI 10.37807/GBMF13806. The work of P.E.A. and G.J. was supported by the U.S. NSF through awards CBET-2525868 and CBET-0932449.
\\

\noindent
\textbf{Author contributions.}
G.J. and L.T. performed the experiments and analyzed the data.  
L.T. performed simulations.
L.T., G.J., J.S.G., and P.E.A. conceived the project, designed the research, and wrote the paper. 
J.S.G. and P.E.A. acquired funding and supervised the project.
\\

\noindent
\textbf{Competing interests.}
The authors declare no competing interests.

\clearpage

\clearpage
\section*{Methods}\label{sec11}


\subsection*{DNA sample preparation}
The polymeric constituents used in this study were double-stranded $\lambda$-DNA (Roche Diagnostics GmbH, Mannheim, Germany) and T4-GT7 DNA (Nippon Gene, Tokyo, Japan). The known contour lengths ($L_c$) of unlabelled molecules are \SI{16.5}{\micro\meter} for $\lambda$-DNA ($M_w = 3.2 \times 10^7$~Da) and \SI{55.6}{\micro\meter} for T4-GT7 DNA ($M_w = 1.0 \times 10^8$~Da). Both DNA species are considered semi-flexible, as their persistence length ($l_p = \SI{50}{\nano\meter}$) is significantly smaller than their total contour length.

Prior to experiments, DNA stock solutions were diluted into a Tris-EDTA buffer and heated to \SI{65}{\degreeCelsius} for 10 minutes. For single-molecule imaging, a small fraction (parts-per-billion) of the molecules was labeled with YOYO-1 fluorescent dye (Invitrogen\texttrademark) at a ratio of 1 dye molecule to 10 base pairs, followed by immediate snap-cooling in an ice bath to achieve a reproducible initial conformational state. This intercalation procedure increases the contour length by approximately $3\%$ \cite{Kundukad2014}. Labeled DNA suspensions were stored at room temperature for 1 hour in the dark to ensure stable dye intercalation before further processing.

The labeled solution was then mixed with unlabeled DNA and added to a viscous Newtonian solvent (92\% w/w glycerol in DI water). To suppress photobleaching and photocleavage during high-intensity imaging, beta-mercaptoethanol (2\% v/v) was added to the final solutions. DNA suspensions were prepared at final normalized molecular concentrations of $c = c^*/8$ or $c = c^*$
Here, $c^*$ is the overlap concentration calculated from $c^* = 3M_w / (4\pi N_A R_0^3)$ \cite{GLopez2024}, where $N_A$ is Avagadro's number and $R_0 = \SIrange[range-phrase=-]{0.7}{1.5}{\micro\meter}$ is the molecule radius of gyration at equilibrium estimated from the $R_0 \propto L_c^{0.61}$ scaling established for flexible double-stranded DNA \cite{Smith1996}. 
For velocity field characterization, the suspensions were seeded with \SI{1}{\micro\meter} diameter polystyrene tracer particles.



\subsection*{Microfluidics and flow characterization}
Microfluidic cross-slot devices ($\SI{100}{\micro\meter}$ width, $\SI{100}{\micro\meter}$ height) were fabricated through standard soft lithography techniques \cite{SoftLito}. Microfluidic channels were cast with polydimethylsiloxane (Dow Corning, SYLGARD 184) and plasma bonded to standard glass microscope slides. DNA suspensions were injected into cross-channel devices using a low-noise syringe pump (Harvard Apparatus) at various flow rates. Imaging of tracer particles and DNA was performed at the mid-height of the channel using bright field and epi-fluorescence imaging, respectively (100$\times$, 1.4 NA). Images were acquired using an intensified high-speed CMOS camera (Photron FASTCAM SA1.1, Hamamatsu high-speed image intensifier) at frame rates of 250 frames per second for flow tracers and ranging from 60 to 250 frames per second for fluorescent molecules. A total of 30,000 frames were acquired per experiment, yielding approximately 15,000 molecule trajectories for the lowest Wi and 3,000 for the highest Wi.

Limiting analysis to steady flow regimes, the velocity field, $\mathbf{u}$, was measured from tracer particles using particle tracking velocimetry \cite{Devasenathipathy2002}. 
The flow-type parameter, $\Phi$, characterized the local fluid deformation in mixed flows and was calculated as $\Phi = (|\dot{\boldsymbol\epsilon}| - |\boldsymbol\Omega|) / (|\dot{\boldsymbol\epsilon}| + |\boldsymbol\Omega|)$. Here, the rate of strain and vorticity tensors are $\dot{\boldsymbol\epsilon} = \tfrac{1}{2}(\nabla \mathbf{u} + \nabla \mathbf{u}^\intercal)$ and $\boldsymbol\Omega = \tfrac{1}{2}(\nabla \mathbf{u} - \nabla \mathbf{u}^\intercal)$, respectively.
The magnitudes of the strain-rate and vorticity tensors are defined as 
$|\dot{\boldsymbol\epsilon}| = \sqrt{\tfrac{1}{2}\dot{\boldsymbol\epsilon}:\dot{\boldsymbol\epsilon}}$ and 
$|\boldsymbol\Omega| = \sqrt{\tfrac{1}{2}\boldsymbol\Omega:\boldsymbol\Omega}$, respectively. 



\subsection*{Image analysis of DNA}
Fluorescence movies of labelled DNA were processed  (MATLAB, Version R2023a) using a custom pipeline designed to maximize the signal-to-noise ratio, while strictly preserving quantitative intensity information \cite{Zhang2019}. 
For each experiment, a time-averaged background image was first generated from the entire stack and subtracted from each frame.
The resulting frames were normalized to a fixed contrast range and subjected to a frame-wise coarse Gaussian blur subtraction (30-pixel kernel) to remove out-of-focus instantaneous contributions.
Residual shot noise was reduced by applying an Anscombe variance-stabilizing transform \cite{Anscombe1948} followed by non-local means filtering \cite{Buades}. The resulting image was binarized, and DNA molecules were identified as objects with connected regions above a predefined area threshold. 
Binary masks were expanded by 10 pixels through morphological dilation and subsequently tracked across frames using a predictive particle tracking algorithm \cite{Ouellette2005} on molecules' centers of mass. 

Once the molecular trajectories were established, the analysis reverted to the background-subtracted raw intensity field to perform quantitative conformational measurements. The instantaneous radius of gyration tensor, $\mathbf{G}_k(t)$, was measured for each molecule, $k$, at time, $t$, from the image intensity as:
\begin{equation*}
\mathbf{G}_k(t) = \frac{\sum_{p, q} I(p, q) ~ \mathbf{x}(p,q) \mathbf{x}^\intercal(p,q)}{\sum_{p, q} I(p, q)},
\end{equation*}
where $I(p,q)$ is the intensity at pixel $(p,q)$ and $\mathbf{x}(p,q)$ is the position vector of pixel $(p,q)$ relative to the molecule's center of mass.



\subsection*{Rheology of DNA suspensions}
The shear viscosity of the DNA suspensions was 0.25$\pm$\SI{0.02}{\pascal}-s as measured on a stress-controlled rheometer (TA Instruments, DH3) with no appreciable normal stresses. 
To characterize their viscoelastic properties, the relaxation times of the DNA suspensions were determined using a capillary breakup extensional rheometer (CaBER) \cite{Anna2001}. 
A $\SI{25}{\uL}$ droplet was placed between two circular cylindrical dowels (radius, $r_0=\SI{2}{\milli\meter}$; initial gap, $h_0=r_0$).  
The gap was rapidly expanded to a final separation, $h_f=4r_0$, within $\Delta t=\SI{0.05}{\second}$, and the evolution of the fluid filament radius, $r(t)$, was recorded at the mid-point using high-speed imaging (Chronos 1.4, Kron Technologies; 200 frames per second). The relaxation time, $\tau$, was determined from the exponential decay of the filament radius in the elasto-capillary thinning regime as $\mathrm{d}[\ln(r(t))]/\mathrm{d}t = -1/3\tau$ \cite{Anna2001}.
Relaxation times were averaged over three independently prepared DNA suspensions with six trials each, resulting in $\tau_{\lambda-\text{DNA}}({c=c^*})=0.07\pm0.01~\text{s}$, $\tau_{\text{T4-GT7}}({c=c^*/8})=0.07\pm0.01~\text{s}$ and $\tau_{\text{T4-GT7}}({c=c^*})=0.25\pm0.05~\text{s}$ for $\lambda$-DNA and T4-GT7 DNA, respectively.
These bulk DNA suspension relaxation times are noted to be significantly smaller than the theoretical molecular relaxation time, defined as $\tau_p=\eta R_0^3/k_BT$, where $\eta = 215~\text{mPa}\cdot \text{s}^{-1}$ is the solvent viscosity, $T$ is absolute temperature ($\SI{293}{\kelvin}$), and $k_B$ is Boltzmann's constant. 
The theoretical molecular relaxation times were predicted to be $\tau_{\lambda-\text{DNA},p}=\SI{20}{\second}$ and $\tau_{\text{T4-GT7},p}=\SI{180}{\second}$.



\subsection*{Numerical simulation of rheological models}

Measured molecular average radius of gyration tensors $\mathbf{G}$ were compared to standard constitutive models \cite{Stone2023}. While these models formally describe the evolution of the conformation tensor $\mathbf{A} \equiv \langle \mathbf{R}\mathbf{R}^\top \rangle$, along the longitudinal axis $\mathbf{G}$ and $\mathbf{A}$ remain proportional with a near-constant factor throughout the Lagrangian trajectories considered here (see Supplementary Information). Here, the evolution of the conformation tensor along Lagrangian fluid trajectories is governed by the upper-convected Maxwell equation:
\begin{equation}
\overset{\nabla}{\mathbf{A}} + \frac{1}{\tau} \mathcal{F}(\mathbf{A}) = \mathbf{0},
\label{eq:conformationEvolution}
\end{equation}
where $\overset{\nabla}{\mathbf{A}} = \frac{\partial \mathbf{A}}{\partial t} + (\mathbf{u} \cdot {\nabla})\mathbf{A} - ({\nabla}\mathbf{u})^\intercal \cdot \mathbf{A} - \mathbf{A} \cdot ({\nabla}\mathbf{u})$ is the upper-convected derivative, $\mathbf{u}$ is the experimental velocity field, and $\tau$ is the molecular relaxation time. 
$\mathcal{F}$ corresponds to an elastic restoring force (see below).

\paragraph*{Numerical resolution}
The governing differential equation (\ref{eq:conformationEvolution}) was integrated forward in time using an explicit Euler scheme with a fixed time step ($\Delta t = 10^{-3}~\text{s} \ll \tau$) utilizing the interpolated experimental flow field. Simulations were initialized by seeding $N = 1000$ independent polymer trajectories across the two inlets. To avoid boundary artifacts, initial coordinates were chosen along the horizontal axis where molecular extension was maximized, and the initial conformation $\mathbf{G}(t=0)$ was directly set to the locally measured experimental tensor. 
Following integration, Lagrangian snapshots were binned onto a $37 \times 37$ Eulerian grid to generate time-averaged maps in the steady flow.
Polymer extension, $L_\parallel/L_c$, and orientation, $\theta$, were measured from the eigenvalues and eigenvectors of the tensor, respectively, for direct comparison with experiments.

\paragraph*{Constitutive models}
The specific deformation dynamics in Eq.~\ref{eq:conformationEvolution} are dictated by the elastic restoring force, $\mathcal{F}(\mathbf{A}) = f(\text{tr}(\mathbf{A})) \mathbf{A} - \mathbf{I}$, where $\mathbf{I}$ is the identity tensor. Two standard models were evaluated: the Oldroyd-B model \cite{Oldroyd1950}, which assumes a linear restoring force ($f=1$), and the FENE-P model \cite{Peterlin1966}, which incorporates finite extensibility via $f(\text{tr}(\mathbf{A})) = L_c^2 / (L_c^2 - \text{tr}(\mathbf{A}))$.

\paragraph*{Anisotropic two-dumbbell method}
To capture the effects of anisotropy in polymer relaxation time (Fig.~\ref{fig4}a-d), an ``anisotropic dumbbell'' method was additionally implemented. In this approach, the local conformation tensor is explicitly represented by two orthogonal dumbbells aligned with its principal axes, which are assigned distinct relaxation times ($\tau_\parallel$ and $\tau_\perp$). At each time step, the individual dumbbell extensions evolve independently under the local flow, subject to a finitely extensible nonlinear elastic (FENE-P) restoring force to prevent non-physical infinite extension, before the intermediate conformation tensor is reconstructed and rediagonalized. This continuous resampling enforces orthogonality at all times. 
Crucially, the frequent, discrete re-diagonalization does not introduce numerical artifacts. 
Provided the integration time step is sufficiently small relative to the polymer relaxation and local flow time scales, the macroscopic tensor evolution converges to a well-defined continuum limit and remains strictly independent of the chosen $\Delta t$.



\subsection*{WLC simulations}
The micro-dynamics of the DNA molecules were modelled using a coarse-grained bead-spring Worm-like Chain (WLC) model.
Each molecule is represented by a chain of $N=40$ beads connected by elastic links of maximum extension, $l = L_c/(N-1)$, with $L_c$ the DNA contour length. 
The time evolution of the $j$-th bead position, $\mathbf{r}_j$, is determined by the Langevin equation:
\begin{align*}
\mathbf{r}_j(t+\Delta t) &= \mathbf{r}_j(t) \\
&+ \mathbf{u}(\mathbf{r}_j)\Delta t + \frac{1}{\zeta} (\mathbf{F}_j^{\text{sp}}-\mathbf{F}_{j-1}^{\text{sp}}) \Delta t + \Delta \mathbf{r}_j^B,
\end{align*}
where the right hand side corresponds to the current position, hydrodynamic advection from velocity field, $\mathbf{u}$, the sum of elastic spring forces, and stochastic Brownian displacement at time, $t$, respectively.
Bead-bead hydrodynamic interactions are ignored, and the effective bead drag coefficient is set to $\zeta \approx 1.12~k_BT$ (estimated from \cite{Larson1999} to match experimental viscosity). 
Fluid advection is determined by linear interpolation of the experimentally measured flow field at the bead center, $\mathbf{u}(\mathbf{r}_j)$.
The tension of the $j$-th spring is determined by its instantaneous length, $l_j = |\mathbf{r}_{j+1}-\mathbf{r}_j|$, using the Marko-Siggia \cite{Marko1995} interpolation formula:
\begin{equation*}
F^{sp}_j = \frac{k_BT}{l_p^\text{eff} } \left[ \frac{1}{4}\left(1-\frac{l_j}{l}\right)^{-2} - \frac{1}{4} + \frac{l_j}{l} \right],
\end{equation*}
which acts along the spring vector with $l_p^\text{eff} = \SI{0.061}{\micro\meter}$ as the effective persistence length \cite{Larson1999}.
The Brownian force is a stochastic term introduced as white noise with zero mean and standard deviation, $\Delta \mathbf{r}_j^B = \sqrt{4 k_B T / \zeta}~\mathcal{U}_j(-1, 1)$, where $\mathcal{U}_j(-1, 1)$ is an independent random variable drawn from a uniform distribution between $-1$ and $1$.

Custom WLC simulations were implemented (MATLAB, Version R2023a) with 1000 molecules simulated per run, and 10 independent runs in total. The time step was chosen as $\Delta t = 10^{-6}~\text{s}$, which was required to prevent numerical instability when evaluating the stiff, nonlinear diverging spring forces near maximum extension, and to accurately integrate the rapid Brownian fluctuations. To obtain realistic initial conformations, uniformly distributed polymer chains were first simulated in a Poiseuille flow corresponding to the microfluidic channel inlet for $10^3~\text{s}$ with steady-state elongation obtained after $10^2~\text{s}$. 
Subsequently, polymer dynamics were simulated in the experimentally measured microfluidic cross-channel flow until advected out of the domain. 
The WLC radius of gyration for chain $k$ was measured as  $\mathbf{G}_k = \frac{1}{N}\sum_{j=1}^{N}\mathbf{x}_j\mathbf{x}_{j}^\intercal$, where $\mathbf{x}_j$ the position vector of bead $j$ relative to the chain's center of mass. 
Similar to experiments, we define the average molecular extension, $L_\parallel$, from the maximum eigenvalue using the convention as $L_\parallel=\sqrt{12\lambda_{\max}}$, where $\lambda_{\max}$ the maximum eigenvalue of the radius of gyration tensor $\mathbf{G}$.


\subsection*{Residual deformation parameter}
To quantify the anisotropic polymer relaxation dynamics (Fig.~\ref{fig4}), we compute the residual deformation, $S_{\parallel,\perp}(t)$, by comparing the measured molecular extension to the theoretical deformation imposed by the local flow field. 
In a purely-extensional, 2D flow with strain rates $\dot{\epsilon}_{xx}$ and $\dot{\epsilon}_{yy}$, the evolution of the extension $L_\parallel$ along the major axis of a molecule oriented at an angle $\theta$ (in the $x$-$y$ laboratory frame) is governed by:
\begin{equation*}
\frac{dL_\parallel}{dt} = \left[ \dot{\epsilon}_{xx}(t) \cos^2\theta(t) + \dot{\epsilon}_{yy}(t) \sin^2\theta(t) \right] L_\parallel - \frac{1}{\tau_\parallel} L_\parallel,
\end{equation*}
where $\tau_\parallel$ represents the relaxation time associated with that specific axis. 
To isolate the relaxation component from the experimental data, we define the quantity $S_\parallel(t)$ as the difference between the logarithm of the measured extension and the cumulative flow-induced strain:
\begin{align*}
S_\parallel(t) &= \ln L_\parallel(t)\\
&- \int_{0}^{t} \left[ \dot{\epsilon}_{xx}(t') \cos^2\theta(t') + \dot{\epsilon}_{yy}(t') \sin^2\theta(t') \right] dt' \\
\label{eq:St}
\end{align*}
where $L_\parallel(t= 0)$ is the initial extension at the start of the Lagrangian trajectory. 
For a molecule exhibiting exponential relaxation, $S_\parallel(t)$ is expected to evolve as:
$S_\parallel(t) = -{t}/{\tau_\parallel}$. 
This analysis is performed independently for the longitudinal ($\parallel$) and transverse ($\perp$) axes of the polymers' radius of gyration tensor. 
The relaxation time for each axis is then determined by a linear fit to $S_{\parallel,\perp}(t)$ versus time. 
This analysis allows us to resolve the disparate timescales governing longitudinal unfolding and transverse compression near the hyperbolic point.    


\bibliography{bibliography}


\clearpage
\renewcommand{\thefigure}{S\arabic{figure}}
\renewcommand{\theequation}{S\arabic{equation}}
\setcounter{page}{1}
\setcounter{figure}{0}
\setcounter{equation}{0}

\nolinenumbers

\section*{Supplementary Information}

\subsection*{Movie caption}
\textbf{Supplementary Video 1. Fluorescence microscopy and radius of gyration detection.}
Time-lapse imaging of representative T4-DNA molecules. The video first displays the raw fluorescence microscopy data, followed by the processed sequence demonstrating the automated radius of gyration detection. Experimental conditions similar to Fig.~\ref{fig1}\textbf{c-d}: concentration $c = c^*/8$ and $\text{Wi} = 0.12$.

\subsection*{Velocity fields}
\begin{figure*}[h]
\centering
\includegraphics[width=2\columnwidth]{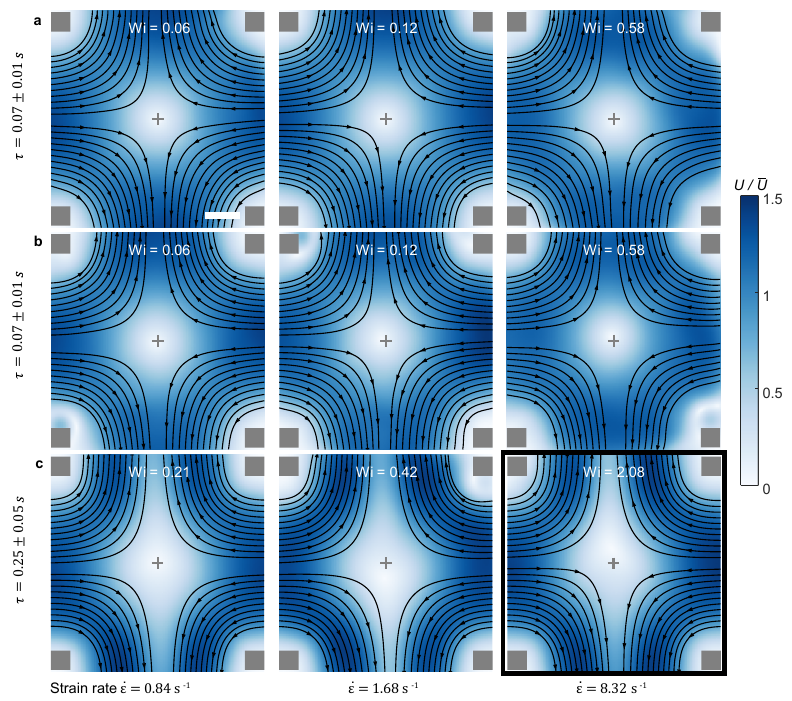}
\caption{\textbf{Effect of polymer concentration and contour length on flow kinematics across Weissenberg numbers.}
Time-averaged, normalized flow speed fields with corresponding streamlines for each tested fluid: (\textbf{a}) semi-dilute $\lambda$-DNA ($c = c^*$), (\textbf{b}) dilute T4-DNA ($c = c^*/8$), and (\textbf{c}) semi-dilute T4-DNA ($c = c^*$).
The fluids in  (\textbf{a-b}) maintain Newtonian flow kinematics regardless of $\text{Wi}$, while (\textbf{c}) exhibits noticeably non-Newtonian flow behavior for all Wi. The framed panel (T4-DNA, $c = c^*$, Wi = 2.08) corresponds to the conditions analyzed in Fig.~\ref{fig3}. Scale bar, $\SI{20}{\micro\meter}$.}
\label{fig:SI_velocityFields}
\end{figure*}

Figure~\ref{fig:SI_velocityFields} shows flow speed maps and corresponding streamlines for the different molecular sizes, concentrations, and Weissenberg numbers investigated (Fig.~\ref{fig2}, main text).
Only T4-DNA ($c = c^*$) exhibits noticeably non-Newtonian flow behavior with extended low-velocity regions downstream of the hyperbolic point, which occur even at low $\text{Wi}$. For all conditions, the velocity field topologies are relatively independent of $\text{Wi}$, for the range of parameters tested here.

\subsection*{Polymer orientation maps}
\begin{figure*}[h]
\centering
\includegraphics[width=2\columnwidth]{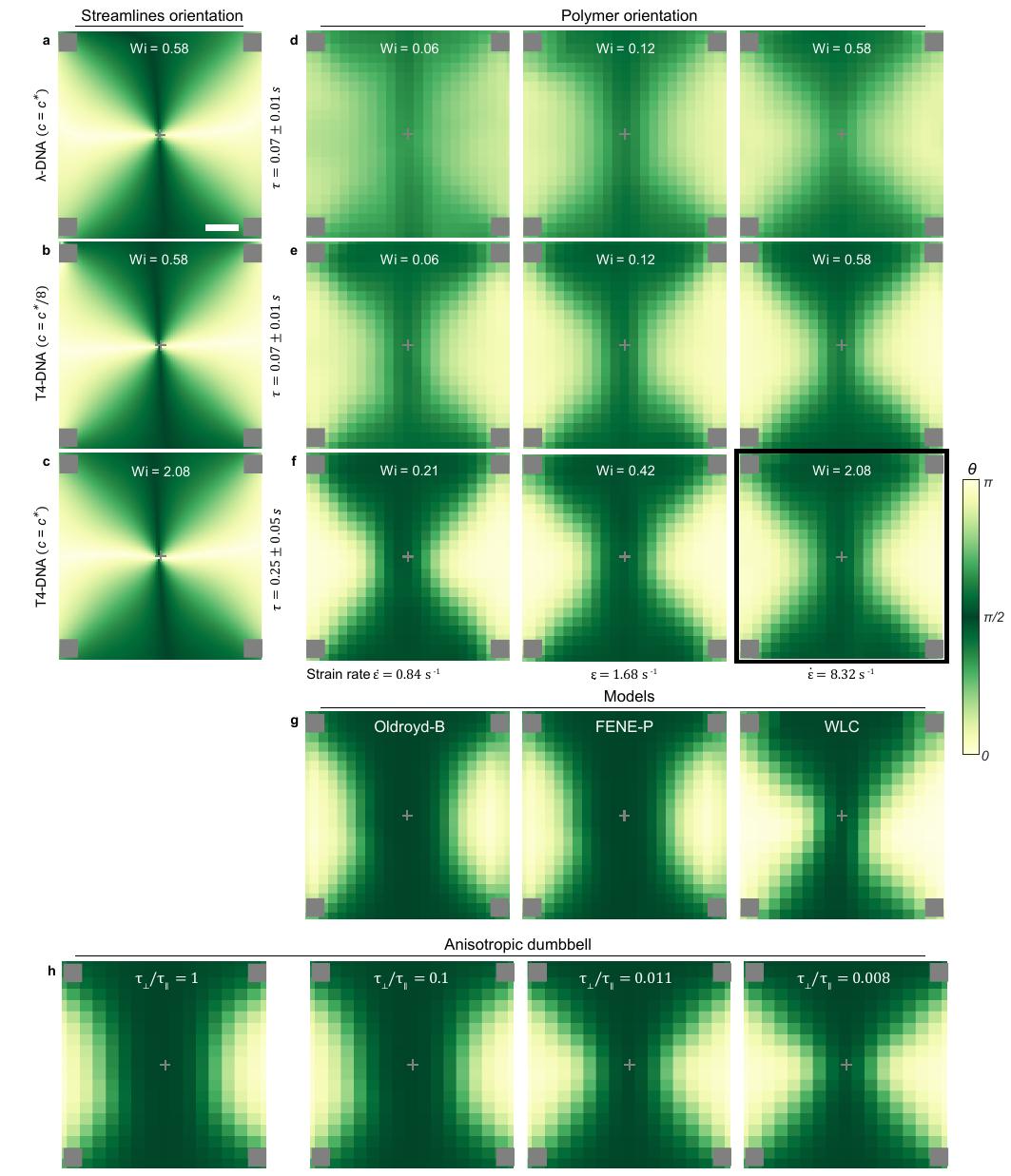}
\caption{\textbf{Effect of polymer concentration and contour length on polymer orientation relative to streamlines across Weissenberg numbers.}
\textbf{a--c.} Streamline orientation of the time-averaged velocity field at the highest Wi investigated for each fluid. Scale bar, \SI{20}{\micro\meter}.
\textbf{d--f.} Spatially-resolved, ensemble-averaged Eulerian polymer orientation maps ($\theta$) determined from the Eulerian average molecule conformation tensor, for the same molecules, concentrations, and strain rates as in Fig.~\ref{fig2} of the main text. The framed panel ($c = c^*$, Wi = 2.08) corresponds to the conditions analyzed in Fig.~\ref{fig3}.
\textbf{g.} Eulerian orientation maps for constitutive and WLC models corresponding to the conditions of Fig.~\ref{fig3} with $\text{Wi} = 2.08$.
\textbf{h.} Eulerian orientation maps for anisotropic dumbbell simulations corresponding to the relaxation time ratios of Fig.~\ref{fig4}e.}
\label{fig:SI_orientationMaps}
\end{figure*}

In addition to the molecular extension fields reported in the main text (Figs.~\ref{fig2}~and~\ref{fig3}), our single-molecule Lagrangian tracking gives access to the ensemble-averaged polymer orientation, $\theta$.
For each Eulerian grid cell, the orientation is defined as the angle of the mean eigenvector corresponding to the maximum eigenvector of the radius of gyration tensor, $\mathbf{G}$, relative to the compressional ($x$-axis) in the laboratory frame. 
Figure~\ref{fig:SI_orientationMaps} reports these orientation maps alongside the corresponding streamline orientation of the underlying flow.
Orientation maps are shown for the same molecules, concentrations, and Weissenberg numbers shown in Fig.~\ref{fig2}, as well as the constitutive and Worm-like Chain (WLC) models in Fig.~\ref{fig3}, and anisotropic dumbbells in Fig.~\ref{fig4}\textbf{e}.
For all conditions, the polymer orientation field closely tracks the underlying streamline geometry across the full range of $\text{Wi}$, except for noticeable deviations near to the hyperbolic point.

\subsection*{Peterlin numerical resolution}

The disagreement in measured extension between experiments and dumbbell models described in Fig.~\ref{fig3} of the main text is consistent with suggestions in prior theoretical work~\cite{Keunings1997}: the pre-averaging closure underlying FENE-P systematically misrepresents transient extensional flow, forcing the configuration distribution to remain Gaussian even as the true FENE distribution localizes near the extensibility limit. However, this closure failure alone does not account for the discrepancy with experiments observed here. 

\begin{figure}[h]
\centering
\includegraphics[width=\columnwidth]{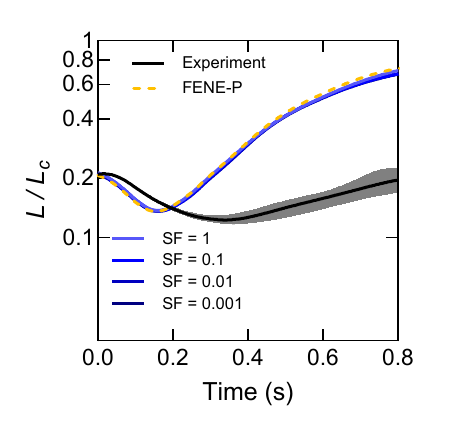}
\caption{\textbf{Assessment of finite-size effects via \mbox{Peterlin} scaling.} Lagrangian evolution of the polymer fractional extension $L_\parallel/L_c$ as a function of time along the representative streamline shown in Fig.~3 of the main text. The experimental measurement is compared with the FENE-P model prediction and ensemble-averaged Peterlin simulations using various polymer scaling factors ($\text{SF} = [1,~0.1,~0.01,~0.001]$). The numerical Peterlin results across all $\text{SF}$ do not deviate appreciably from the FENE-P prediction.}
\label{fig:Peterlin_SF}
\end{figure}

To verify our numerical approach for the FENE-P model and to specifically investigate whether finite molecular size in experiments impact the discrepancy with models, we implemented a Peterlin approach \cite{Keunings1997}. 
In this method, the macroscopic conformation tensor is represented by an ensemble of dumbbells at each inlet point, which randomly sample the experimental conformation tensor.
Finite extensibility is imposed individually on each dumbbell, which avoids the ensemble-level Gaussian closure entirely (Fig.~\ref{fig:Peterlin_SF}).
We find that the resulting dynamics recover the standard FENE-P prediction, indicating that the discrepancy with experiments is architectural rather than an artifact of the Gaussian closure.

\paragraph{Statistical initialization} 
The initialization procedure, implemented via a geometric construction algorithm, generates a representative ensemble of dumbbells that statistically reproduces the measured radius of gyration tensor, $\mathbf{G}_{\text{exp}}$. For each inlet location, $\mathbf{x}_0$, the procedure follows a three-step reconstruction:

\begin{enumerate}
\item \textbf{Tensor decomposition:} The local experimental radius of gyration tensor, $\mathbf{G}_{\text{exp}}$, is diagonalized to identify its principal axes (eigenvectors $\mathbf{v}_1$, $\mathbf{v}_2$) and principal radii of gyration (eigenvalues $\lambda_1$, $\lambda_2$).

\item \textbf{Elliptical mapping:} An elliptical boundary is defined in the local frame with semi-major and semi-minor axes $a = \sqrt{2\lambda_1}$ and $b = \sqrt{2\lambda_2}$. This strict scaling ensures the generated dumbbell ensemble exactly recovers the magnitude of the experimental tensor.

\item \textbf{Stochastic sampling:} An ensemble of $10$ independent dumbbells is generated at each inlet coordinate. Rather than sampling physical orientations uniformly, a parametric angle $\theta \in [0, 2\pi)$ is sampled uniformly from a unit circle. The unit vector $\mathbf{u} = [\cos\theta, \sin\theta]^\top$ is then mapped onto the ellipse surface via an affine transformation using the semi-axes and principal eigenvectors. This parametric mapping naturally concentrates the physical spatial distribution toward the major axis. The resulting vectors define the relative bead positions $\mathbf{r}_k$, placing the two beads of the $k$-th dumbbell at spatial coordinates $\mathbf{x}_0 \pm \mathbf{r}_k$.
\end{enumerate}


\paragraph{Dynamic evolution} Following initialization, the ensemble of dumbbells is evolved in time using a Lagrangian predictor-corrector scheme. The motion of each bead $j \in \{1,2\}$ associated with dumbbell $k$, at position $\mathbf{r}_{k;j}$, is governed by the balance of hydrodynamic drag and non-linear elastic forces:
\begin{equation}
\mathbf{r}_{k;j}(t+\Delta t) = \mathbf{r}_{k;j}(t) + \mathbf{u}(\mathbf{r}_{k;j})\Delta t + \mathbf{F}_{k;j}^{\text{spring}}\Delta t,
\end{equation}
where the flow velocity $\mathbf{u}(\mathbf{r}_{k;j})$ is sampled directly at the instantaneous position of each bead using linear interpolation of the experimental flow field (Fig.~2, main text). The elastic response of each dumbbell is modeled using the FENE-P force law to account for the finite extensibility of the DNA molecule:
\begin{equation}
\mathbf{F}_{k;j}^{\text{spring}} = (-1)^{j-1}\,\mathbf{e}_k\,\frac{1}{4\tau} \frac{L_k}{1-(L_k/L_c)^2},
\end{equation}
which acts along the dumbbell connector vector, drawing the two beads together, with $L_k=\|\mathbf{r}_{k;2}-\mathbf{r}_{k;1}\|$ the instantaneous length of dumbbell $k$, and $\mathbf{e}_k = ({\mathbf{r}_{k;2}-\mathbf{r}_{k;1}})/{\|\mathbf{r}_{k;2}-\mathbf{r}_{k;1}\|}$ the dumbbell unit vector.

\paragraph{Finite size effects} To verify whether the finite size of the molecules, and thus the spatial variation of the velocity gradient across the polymers, contributes to deviations from the continuous FENE-P model, we computed the elongation evolution along the streamlines (Fig.~3, main text) for different scaling factors, $\text{SF}$, of the dumbbell size (Fig.~\ref{fig:Peterlin_SF}). 
The ensemble-averaged elongation obtained from the Peterlin scheme matches the continuous FENE-P constitutive equation exactly, regardless of the scaling factor in the range $10^{-3} \le \text{SF} \le 1$, where $\text{SF}=1$ corresponds to the experimentally measured mean molecular extension. 
This collapse of the numerical curves confirms that the physical dimensions of the polymer relative to the flow gradients (i.e. finite size effects) do not account for the discrepancy between model and experiment.

\subsection*{Equivalence of longitudinal extension for radius of gyration and conformation tensors}

The radius of gyration tensor, $\mathbf{G}$, measured experimentally, and the end-to-end conformation tensor, $\mathbf{A} = \langle \mathbf{R}\mathbf{R}^\top\rangle$, evolved by classical constitutive models are proportional at equilibrium for Gaussian chains \cite{Bird1987} but diverge when out of equilibrium. To verify that the longitudinal comparison underlying Fig.~3 of the main text is robust to this distinction, we compute both extractions from the same WLC simulation along the same Lagrangian trajectories.
Specifically, for the radius of gyration tensor of each chain, we have $L_\parallel^G = \sqrt{12\,\lambda_{\max}(\mathbf{G})}$, and from the chain's end-to-end vector we have $L_\parallel^A = |\mathbf{R}|$.

\begin{figure*}[h]
\centering
\includegraphics[width=2\columnwidth]{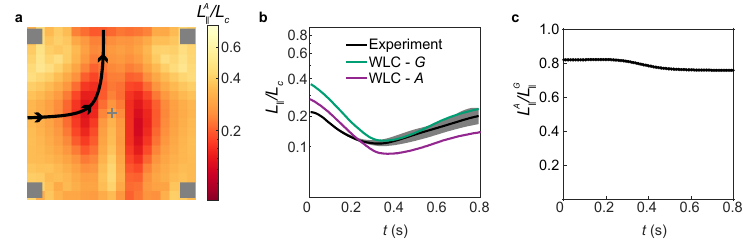}
\caption{\textbf{Comparison of longitudinal extension extracted from the radius of gyration and conformation tensors in WLC simulations.} 
\textbf{a.} Spatially-resolved Eulerian map of $L_{\parallel}^{A}/L_c$ from the WLC simulation of T4-DNA at the conditions of Fig.~3a of the main text, with the representative streamline for Lagrangian analysis overlaid in black. Grey cross and squares are the hyperbolic stagnation point and channel corners, respectively. 
\textbf{b.} Lagrangian evolution of $L_{\parallel}/L_c$ along the streamline in \textbf{a} for the experimental measurement (black; shaded region is the standard deviation across nearby streamlines). 
For WLC simulations, the fractional extension is shown based on both the radius of gyration tensor ($L_{\parallel}^{G} = \sqrt{12\,\lambda_{\max}(\mathbf{G})}$; green) and the conformation tensor ($L_{\parallel}^{A} = |\mathbf{R}|$; magenta).
\textbf{c.} Ratio of the conformation tensor to the radius of gyration tensor extension along the trajectory shown, exhibiting a near-constant value of approximately $L_{\parallel}^{A}/L_{\parallel}^{G} \approx 0.8$.}
\label{fig:SI_conformation_WLC}
\end{figure*}

The two extractions yield consistent spatial topologies (Fig.~\ref{fig:SI_conformation_WLC}a) and follow identical temporal dynamics along a representative streamline (Fig.~\ref{fig:SI_conformation_WLC}b), with the experimental measurement largely falling between the two WLC curves. 
The ratio between the two WLC extractions remains near-constant at approximately $L_{\parallel}^{A}/L_{\parallel}^{G} \approx 0.8$ throughout the stretch-coil-stretch transition (Fig.~\ref{fig:SI_conformation_WLC}c). Because $L_\parallel^G$ and $L_\parallel^A$ differ by a multiplicative factor -- rather than in their temporal evolution -- the discrepancy between experiment versus FENE-P and Oldroyd-B model predictions reported in Fig.~3 of the main text appears independent of the tensor used for the comparison.

\subsection*{Rouse model sub-chain relaxation time}
The anisotropic FENE-P dumbbell framework represents the polymer conformation using two finitely-extensible, orthogonal dumbbells that deform with the fluid gradient but relax at independent rates, $\tau_{\parallel,\perp}$ (Fig.~\ref{fig4}c). 
For a polymer chain with a longitudinally measured relaxation rate of $\tau_\parallel = \SI{1.25}{\second}$, a parametric exploration yields a best-fit transverse relaxation time of $\tau_\perp \approx 0.011 \tau_\parallel \approx \SI{13}{\milli\second}$ (Fig.~\ref{fig4}e). 
Although smaller than the measured $\tau_\perp$ (Fig.~\ref{fig4}d), 
the physical origin of this rapid transverse timescale is rationalized by the Rouse model \cite{Rouse1953,Doi1986}, where the global relaxation time of a chain with $N \approx 40$ persistence lengths is $\tau_N = \tau_0 N^2 / \pi^2$. 
Equating the measured $\tau_\parallel$ to $\tau_N$ yields an elementary fundamental spring relaxation time of $\tau_0 \approx \SI{8}{\milli\second}$, which is remarkably consistent with the phenomenological best-fit described in the main text.

\backmatter

\end{document}